\documentclass[a4paper,11pt]{article}
\pdfoutput=1 % if your are submitting a pdflatex (i.e. if you have
\usepackage{jinstpub} % for details on the use of the package, please
\title{\boldmath Increasing the efficiency of photon collection in LArTPCs: the ARAPUCA light trap}

\author[a]{G. Cancelo,}
\author[a]{F. Cavanna,}
\author[a]{C. O. Escobar,}
\author[a,b,1]{E. Kemp,\note{Corresponding author.}}
\author[c]{A. A. Machado,}
\author[a]{A. Para,}
\author[b]{E. Segreto,}
\author[e]{D. Totani,}
\author[f]{D. Warner}

\affiliation[a]{Fermilab National Accelerator Laboratory,\\Batavia, 60510 (IL) USA}
\affiliation[b]{Universidade Estadual de Campinas - UNICAMP,\\Campinas, 13083-859 (SP) Brazil}
\affiliation[c]{Universidade Federal do ABC,\\Santo Andr\'{e}, 09210-580 (SP) Brazil}
\affiliation[d]{Universit\`{a} degli Studi dell'Aquila,\\ L'Aquila, 67100 (ABR), Italia}
\affiliation[e]{Colorado State University,\\ Fort Collins, 80523 (CO), USA}

\emailAdd{kemp@ifi.unicamp.br}

\abstract{The Liquid Argon Time Projection Chambers (LArTPCs) are a  choice for the next generation of large neutrino detectors due to their optimal performance in particle tracking and calorimetry. The detection of Argon scintillation light plays a crucial role in the event reconstruction as well as the time reference for non-beam physics such as supernovae neutrino detection and baryon number violation studies. In this contribution, we present the current R\&D work on the ARAPUCA (\textbf{A}rgon \textbf{R}\&D \textbf{A}dvanced \textbf{P}rogram at \textbf{U}NI\textbf{CA}MP), a light trap device to enhance Ar scintillation light collection and thus the overall performance of LArTPCs. The ARAPUCA working principle is based on a suitable combination of dichroic filters and wavelength shifters to achieve a high efficiency in light collection. We discuss the operational principles, the last results of laboratory tests and the application of the ARAPUCA as the alternative photon detection system in the protoDUNE detector.}


\arxivnumber{1802.09726 } % only if you have one

\proceeding{LIDINE 2017: Light Detection in Noble Elements\\
22-24 September 2017\\
SLAC National Accelerator Laboratory}

\begin{document}
\maketitle
\flushbottom

\section{Overview}
\label{sec:intro}

Liquid Argon Time Projection Chambers (LArTPCs) technology are the best choice for the next generation of large neutrino experiments \cite{DUNE, SBN} and frequently has been also an option for Dark Matter detectors \cite{LEPwithNL}. Neutrinos or Dark Matter candidates can interact with argon nuclei producing charged particles as final states. The charged particles cross the liquid argon (LAr) medium promptly producing scintillation light and free electrons by ionization. The scintillation is produced either by the direct excitation of an Ar atom followed by the formation of an excimer which then decays ($Ar^* + Ar \rightarrow Ar_2^* \rightarrow 2Ar + \gamma \; (128nm)$) or by ionization, recombination, and finally de-excitation ($Ar^+ +Ar \rightarrow Ar^+_2 + e^- \rightarrow Ar_2^* \rightarrow 2Ar + \gamma \; (128nm)$).
These characteristic 128 nm photons - VUV photons (Vacuum Ultra-Violet) - are abundantly emitted, about 40,000 photons per MeV of ionization energy loss. Light emission is much faster ($\sim$ ns) than the drift of ionized charges to the anode planes ($\sim \mu$s). Therefore the collection of these photons provides a natural time reference ($t_0$) for the particle interactions inside the TPC and is essential to improve the timing accuracy of the event dynamics and thus the reconstruction of the particles tracks. Moreover, the scintillation light provides the $t_0$ for non-beam events such as supernovae neutrinos and nucleon decay rare signatures. 
The LArTPCs can use both charge and light for trigger purposes and also to improve the detector performance. A fraction of the ionization electrons is not collected due to their capture by electronegative impurities in the LAr (mainly $O_2$). The efficiency of the photon detection for low energy events, such as those of supernovae neutrinos, depends critically on the signal-to-noise ratio. It is extremely advantageous to eliminate background events of $^{39}Ar$ and $^{222}Rn$, which requires a threshold in few photoelectrons (PE). The light might also be used to find the vertex of the interaction improving the energy resolution relevant for supernova neutrinos. Furthermore, typical light sensors for particle detectors based on liquefied noble gases, whether cryogenic phototubes (PMTs) or silicon photomultipliers (SiPMs) generally are not suitable for VUV photons\footnote{There are special devices built for VUV at very high cost.} \hspace{0.3mm} since the quantum efficiencies of these devices \hspace{0.3mm} are optimal only at \hspace{0.3mm} wavelengths $\lambda$ > 300 nm.  Thus, there is a great challenge to developing photon collection systems for LArTPCs that increase the detection efficiency allowing the experiments to fully achieve the scientific goals. This can be done exploring the direct detection of VUV photons or by down-shifting the photon wavelengths before detection. In both cases, higher efficiencies is a strong requirement that can be achieved by a high-performance photon collection device. 

\section{ARAPUCA device: the working principles and concepts}
\label{sec:arapuca}

One proposal for an efficient photon collector is a light trap called ARAPUCA\footnote{Arapuca is a handcrafted trap originally used by the Guarani Brazilian Indians to catch birds, monkeys and other small animals} \cite{Arapuca}. The central concept of ARAPUCA is to capture photons within a box with highly reflective internal surfaces (> 98\% reflectivity).
%, so that the detected photon detection efficiency is high even with a limited active coverage \ footnote {covered area by the SiPMs.} inside the box. 
The key point of the device is a dichroic filter made from a vitreous substrate with acrylic multilayer film. The filter has the property of being highly transparent to wavelengths below a cut ($\lambda_{cut}$) and highly reflective above it. The dichroic filter is the acceptance window for the device. The filter has its outer side coated by a wavelength shifter (WLS). The WLS on the outer side, $W_{out}$, first converts the scintillation light from the LAr to a wavelength $\lambda_1$, with $\lambda_1$ < $\lambda_{cut}$, and the filter is transparent to the photons, allowing them to enter the box. Another WLS layer, $W_{in}$, coats the inner side of the filter, or alternatively, the inner walls of the box. The $W_{in}$ makes a second conversion on the photons to a wavelength $\lambda_2$, with $\lambda_2$ > $\lambda_{cut}$. After the second conversion the filter becomes highly reflective, trapping the photons in the the box. Inside the ARAPUCA there are photosensors viewing the internal volume. So far we have used SiPMs as photosensors, filters with $\lambda_{cut} = 400$nm, P-Terphenyl\footnote{1,4-Diphenylbenzene} (pTp, emmison peak at $\lambda = $ 338 nm) as $W_{out}$, and Tetra-Phenyl-Butadiene\footnote{1,1,4,4-Tetraphenyl-1,3-butadiene} (TPB, emmison peak at $\lambda = $ 420 nm) as $W_{in}$. 
%Figure \ ref {arapuca-principle} illustrates the principle of operation.

\section{The experiment at TallBo cryostat}
\label{sec:TallBoRun}

%The TallBo cryostat (56 cm inner diameter, 183 cm depth). The radioactive source ($^{241}Am$ --  $1 \mu$ Ci -- 5.4 MeV line) position  is controlled from outside. In the source holder is installed a SiPM coated with TPB for trigger purposes. Scintillation light is produced promptly by ionizing particles through the liquid argon.\\

The experiment described in this section was aimed i) to check the performances of basic components of the ARAPUCAs; and ii) obtain a measurement of the absolute efficiency for photon collection. The experiment was performed at TallBo cryostat, in March/2017. TallBo is a cylindrical  cryostat with 56 cm inner diameter and up to a 183 cm liquid depth for a total volume of liquid argon of 451 liters available in the cryogenic facilities at Fermilab. The cryostat is vacuum-jacketed for insulating the liquid argon. The argon is filtered to remove oxygen and water contamination. TallBo has a nitrogen cooled condenser in order to provide a closed system without boil-off of the liquid argon. A vertical mover, 6 feet in length, is available. It consists of a slotted channel and a screw capable of vertically moving objects mounted on it. The screw has a radial distance from the cylinder center $d_r=$12.7 cm. The vertical position $z$ can be controlled from outside by a graduated scale.
The experiment was performed by testing six different configurations of the ARAPUCAs. A summary of the configurations can be found in Table \ref{tab1}. In the configurations with Vikuit\textsuperscript\textregistered 3M Enhanced Specular Reflector\footnote{\url{https://www.3m.com/3M/en_US/company-us/search/?Ntt=ESR&LC=en_US&co=cc&gsaAction=scBR&type=cc}}, the lining was applied only over the bottom of the ARAPUCAs. All other cases the reflectivity on the bottom or on the side walls was provided by the properties of the box materials themselves (PTFE\footnote{Polytetrafluoroethylene .} or Accuflect\textsuperscript\textregistered  Light-Reflecting Ceramic\footnote{\url{http://accuratus.com/accuflprods.html}}). Filters with same technical specifications but from different vendors were also tested to compare their relative performances\footnote{ \scriptsize{* \url{https://www.edmundoptics.com/} ; 
$\dag$ \url{http://www.asahi-spectra.com} ; $\ddag$\url{http://www.omegafilters.com}}.}. In a first approximation, the inner reflectors have no considerable differences in their reflection coefficients, one of the purposes of the test was to assess the ease of machining, handling, and assembly with different materials.  The WLS coatings were made using evaporation technique. The thickness of the coatings is controlled by monitoring the vibrational frequency of a piezoelectric quartz crystal placed inside the vacuum chamber of the evaporator.

\begin{table}[htbp]
\centering
\caption{\label{tab1} Configurations of the ARAPUCAs tested at TallBo cryostat. Configuration 1 used two (2) filter plates to achieve a comparable window area to the others. The wavelength cut-off of all filters is $\lambda_{cut} = $ 400 nm. In first column, there is the channel numbering, relevant for further discussions in this paper, and the position of each ARAPUCAs in the frame (row and column of the frame, see more details in Fig. \ref{TallBoFrame}). The $W_i$ coating (TPB) was applied in two different configurations: over the inner side of the filter plate (denoted by {\it f} in the table) or over the lining reflector placed on the backplane of the ARAPUCA (denoted by {\it l}).}

\begin{center}
 \begin{tabular}{||c c c c c||} 
 \hline
 Configuration & Filter & pTp thickness & TPB thickness & Internal reflector\\
  & (vendor and size) & $\mu$g/cm$^2$ & $\mu$g/cm$^2$ &   lining \\ [0.5ex] 
 \hline\hline
 1: ch0, r1c1 & 2 Edmund\textsuperscript{*} 2.5$\times$3.6 cm$^2$ & 200 & 250 \it f & Teflon\\ 
 \hline
 2: ch1, r2c1 & Asahi\textsuperscript{$\dag$} 5.0$\times$5.0 cm$^2$ & 200 & 250 \it f & Vikuit\\
 \hline
 3: ch3, r2c2  & Omega\textsuperscript{$\ddag$}  5.0$\times$7.0 cm$^2$  & 300 & 250 \it l &   Vikuit \\
 \hline
 4: ch4, r3c2 & Omega 5.0$\times$7.0 cm$^2$ & 500 & 250 \it l & Vikuit \\
 \hline
 5: ch5, r1c3 & Omega 5.0$\times$5.0 cm$^2$ & 300 & 300 \it f & Vikuit \\
 \hline
 6: ch6, r3c3 & Asahi 5.0$\times$5.0 cm$^2$ & 200 & 250 \it f & Accuflect\\ [1ex] 
 \hline
\end{tabular}
\end{center}
\end{table}
%\normalsize                                                                                       

\begin{figure}[htbp]
\centering % \begin{center}/\end{center} takes some additional vertical space
\includegraphics[width=0.9\textwidth]{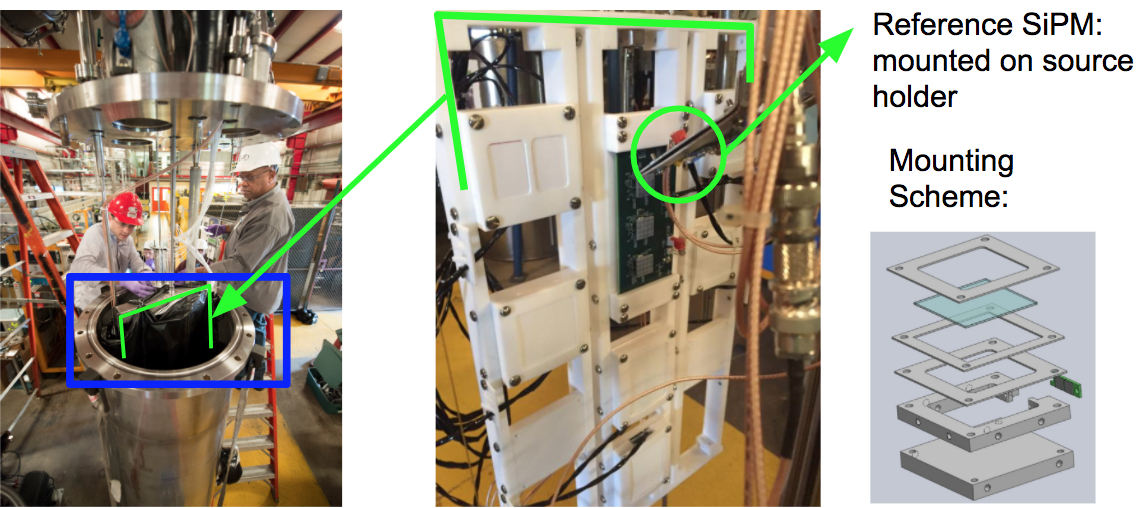}
%\qquad
%\includegraphics[width=.4\textwidth,origin=c,angle=180]{example-image-b}
% "\includegraphics" from the "graphicx" permits to crop (trim+clip)
% and rotate (angle) and image (and much more)
\caption{\label{TallBoFrame} Center panel: The different ARAPUCAs in the rack. The correspondence with the Configurations in Table \ref{tab1} are made by row (r1, r2, r3 from top to bottom) and column (c1, c2, c3 from left to right). The reference SiPM (trigger) is shown in the detail. Left panel: The rack installation inside TallBo. Right panel: exploded view of the ARAPUCA basic design, showing the box base and walls, SiPM board, dichroic filter window, and its fixture frames.}
\end{figure}

The six ARAPUCAs were mounted in a rack\footnote{\label{array}An additional array of bare SiPMs coated with TPB was also mounted in the rack, with the purpose to test a summing pre-amplifier board, but the discussion on this device is out of the scope of this work.} which in turn was placed inside the cryostat. The ARAPUCA's rack was hung by a fixture rod coincident with the cylinder axis. An  alpha source  ($^{241}Am$ --  $1 \mu$ Ci -- 5.4 MeV line) was mounted on the mover. On top of the source holder was installed a SiPM coated with TPB for trigger purposes (reference SiPM).  Figure \ref{TallBoFrame} shows a picture of the rack and its installation in TallBo.   

\begin{figure}[htbp]
\centering % \begin{center}/\end{center} takes some additional vertical space
\includegraphics[width=0.80\textwidth,trim=0 0 0 0,clip]{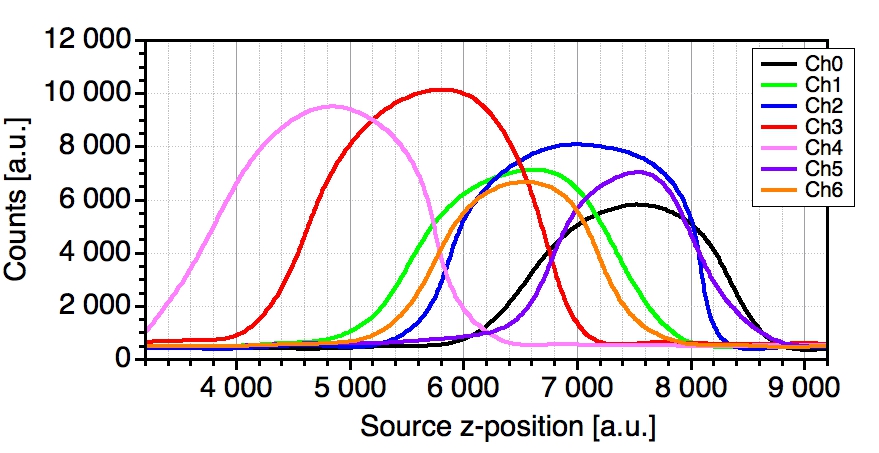} 
%\qquad
%\includegraphics[width=.4\textwidth,origin=c,angle=180]{example-image-b}
% "\includegraphics" from the "graphicx" permits to crop (trim+clip)
% and rotate (angle) and image (and much more)
\caption{\label{ArapucasCounting}  ARAPUCAs counting rate in a common prescaled time for all channels as function of the source position $z$. LAr level is $\sim$ 9200. This plot shows and excellent correspondence with the disposition of the ARAPUCAs in the mounting frame (see Table \ref{tab1} for details on the ARAPUCAs' positions.) Curves are {\it splines} connecting points measured in steps of $z =$200. Channel 2 is the array of SiPMs (see footnote \ref{array}).}
\end{figure}

Each ARAPUCA was instrumented with two 6 $\times$ 6 mm$^2$ Sensl MicroFC-60035-SMT\footnote{\url{http://sensl.com}} SiPMs connected in parallel, soldered in a single board. The experiment was performed by digitizing the waveforms of all ARAPUCAs with a custom SiPM Signal Processor (SSP) described in \cite{SSP}.  The SSP was triggered by signals from the reference SiPM. We took data scanning the ARAPUCAs plane with the source, keeping $d_r=$12.7 cm fixed and at different $z$ positions. Changes in the solid angle $\Omega (z)$ between the source and each ARAPUCA modulate the %intensity of scintillation photons impinging the acceptance windows and so the 
number of detected photons $N_{det} (z)$. Figure \ref{ArapucasCounting} illustrates this effect. The maxima of the  curves indeed correspond to $z$ that maximizes the solid angle viewed by the correspondent ARAPUCA.

%We took data with several values of $z$ including those positions where maxima in counting rates of each Arapuca where observed. ositioned in Specific scanning positions were chosen based on the maximum counting rate  scanning positions $z_i$ were chosen by the alignment of the source with the center of each Arapucas row in the rack (see Figure \ref{TallBoFrame}).

%\textit{\textbf{\textcolor{blue} {excimers$^*$}}} {$\textcolor{blue} \rightarrow$ VUV photons \small (128 nm) + Ar + Ar}
%\\ singlet: $\tau_{fast} \sim$ \textcolor{red}{\normalsize 6 ns} \hspace{2mm} ; \hspace{2mm} triplet: $\tau_{slow} \sim$ \textcolor{red}{ \normalsize 1.6 $\mu$s}

\section{Data Analysis: Efficiency Calculation}
\begin{figure}[htbp]
\centering % \begin{center}/\end{center} takes some additional vertical space
\includegraphics[width=.95\textwidth,trim=0 60 0 80,clip]{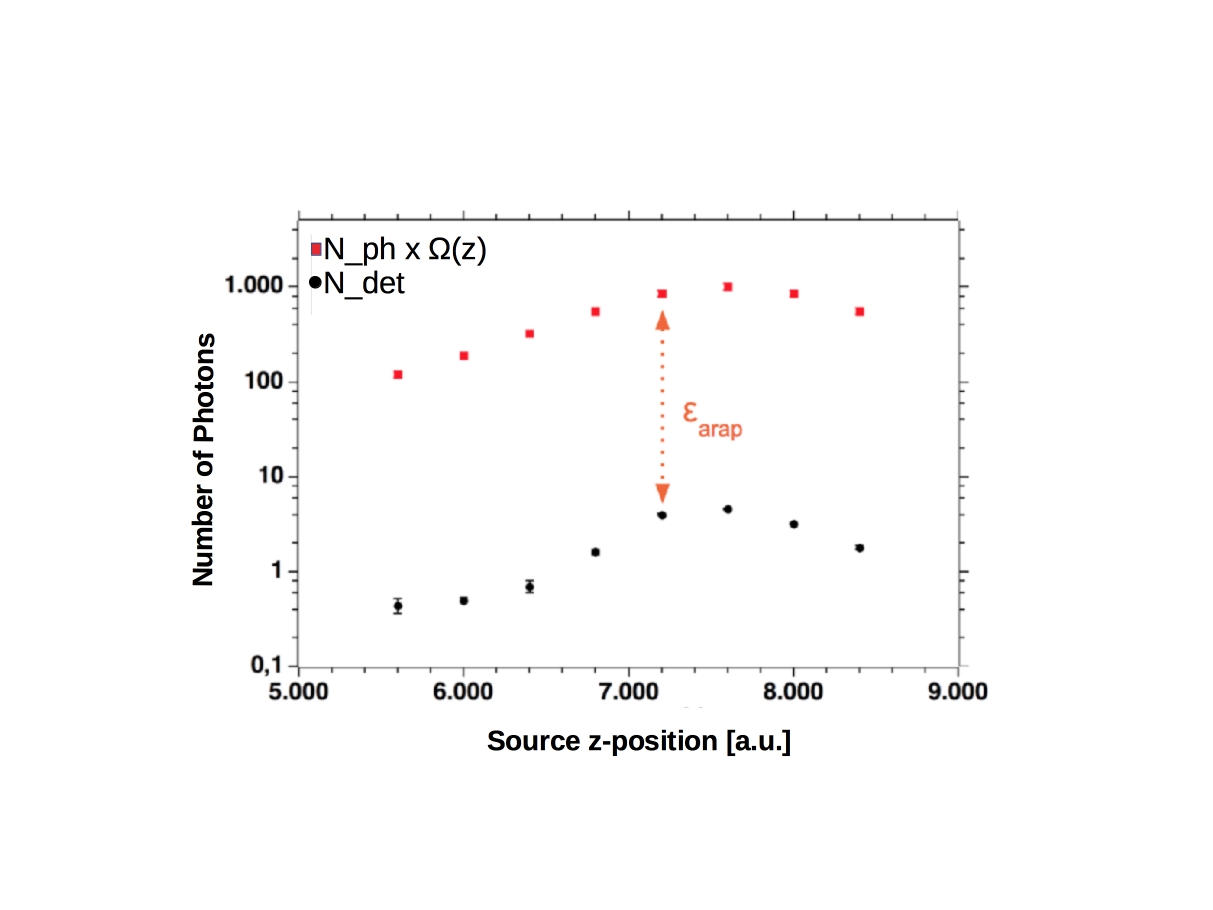}
\caption{\label{PhotoCountxZ} The expected number of photons per $\alpha$-decay in the case $\epsilon_{arap}=1$ (red) and true data (black). The overall scale factor is given by the effective value of $\epsilon_{arap}$.}
\end{figure}

Photo-counting is achieved using as normalization factor the integrated charge of the waveforms from single avalanches in the SiPMs, easy to be identified in dark pulses. Then, for each source position we have the average number of detected photo-electrons $N_{det}(z)$ obtained from the PE number spectrum of each ARAPUCA. $N_{det}(z)$ can be factorized as:  

\begin{equation}
\label{Ndet}
N_{det}(z) = N_{ph} \times \frac{\Omega(z)}{4\pi} \times \epsilon_{arap}
\end{equation}

where $N_{ph}$ is  the expected number of photons produced in each $\alpha$-decay from the source and $\epsilon_{arap}$ is the efficiency of the ARAPUCA. $N_{ph}$ is obtained from $N_{ph} = \frac{<E_{\alpha}>}{W_{ph}} \cdot q$, where $<E_{\alpha}>$ is the average value of the smeared energy\footnote{The $^{241}Am$ $\alpha$-line is smeared by the thin protection gold plate of the source assembly.} spectrum of the $\alpha$-decays from $^{241}Am$. $<E_{\alpha}> =$ 4.1 MeV was estimated by extraction of random values from the experimental $E_\alpha$ distribution found in \cite{DayaBaySpec}; $W_{ph}=19.5$ eV is the average energy to produce a single photon; $q = 0.72\pm 0.04$ is the quenching factor for $\alpha$ particles in LAr \cite{Doke}. Background is taken into account by subtraction from PE spectra of each ARAPUCA the dark spectra obtained from data with $z$ corresponding to the flat part of counting curves of Figure \ref{ArapucasCounting}. Figure \ref{PhotoCountxZ} shows a plot for the ARAPUCA at (ch0, r1c1) with two data sets: the expected number of photons calculated from Eq. \ref{Ndet} in the assumption of $\epsilon_{arap}=1.0$ (red squares), and $N_{det}(z)$ (black dots). All other ARAPUCAs have similar behavior. %The difference between expectations for in the counting is given by the efficiency $\epsilon_{arap}$ which acts as an overall scale factor. 
Since $\epsilon_{arap}$ is the overall scale factor between the two data sets, it can be obtained for each ARAPUCA by $\chi ^2$ minimization between the expected and observed data taking $\epsilon_{arap}$ as free parameter. 

Moreover, for the sake of comparison between different configurations, one can define an effective photo-collection area by the product $A_{eff}=A \times \epsilon_{arap}$, where $A$ is the acceptance window area. Table \ref{results} summarizes the results obtained in the TallBo experiment. 

%The steps described above were repeated for each one of the six Arapucas. Table \ref{results} summarizes the results obtained from the six configurations of the Arapucas in this experiment. 

\begin{table}[htbp]
\centering
\caption{\label{results} Efficiencies of the six ARAPUCAs.}

\begin{center}
 \begin{tabular}{||c c c c||} 
 \hline
 Configuration & $\epsilon_{arap} [\%] $ & $A_{eff}$ [$cm^2$] & \\
[0.5ex] 
 \hline\hline
 1: ch0, r1c1 & $0.45 \pm 0.03$ &  $0.081 \pm 0.003$ &  \\ 
 \hline
 2: ch1, r2c1 & $0.41 \pm 0.03$ &  $0.102 \pm 0.008$ &  \\
 \hline
 3: ch3, r2c2 & $0.33 \pm 0.01$ & $0.116 \pm 0.004$ &  \\
 \hline
 4: ch4, r3c2 & $0.25 \pm 0.02$ & $0.088 \pm 0.004$ &  \\
 \hline
 5: ch5, r1c3 & $0.40 \pm 0.02$ & $0.100 \pm 0.006$ &  \\
 \hline
 6: ch6, r3c3 & $0.40 \pm 0.03$ & $0.100 \pm 0.008$ &  \\ [1ex] 
 \hline
\end{tabular}
\end{center}
\end{table}

\section{ARAPUCAs at protoDUNE}

\begin{figure}
  \centering
  \includegraphics[width=.9\linewidth, trim=0 90 0 93,clip]{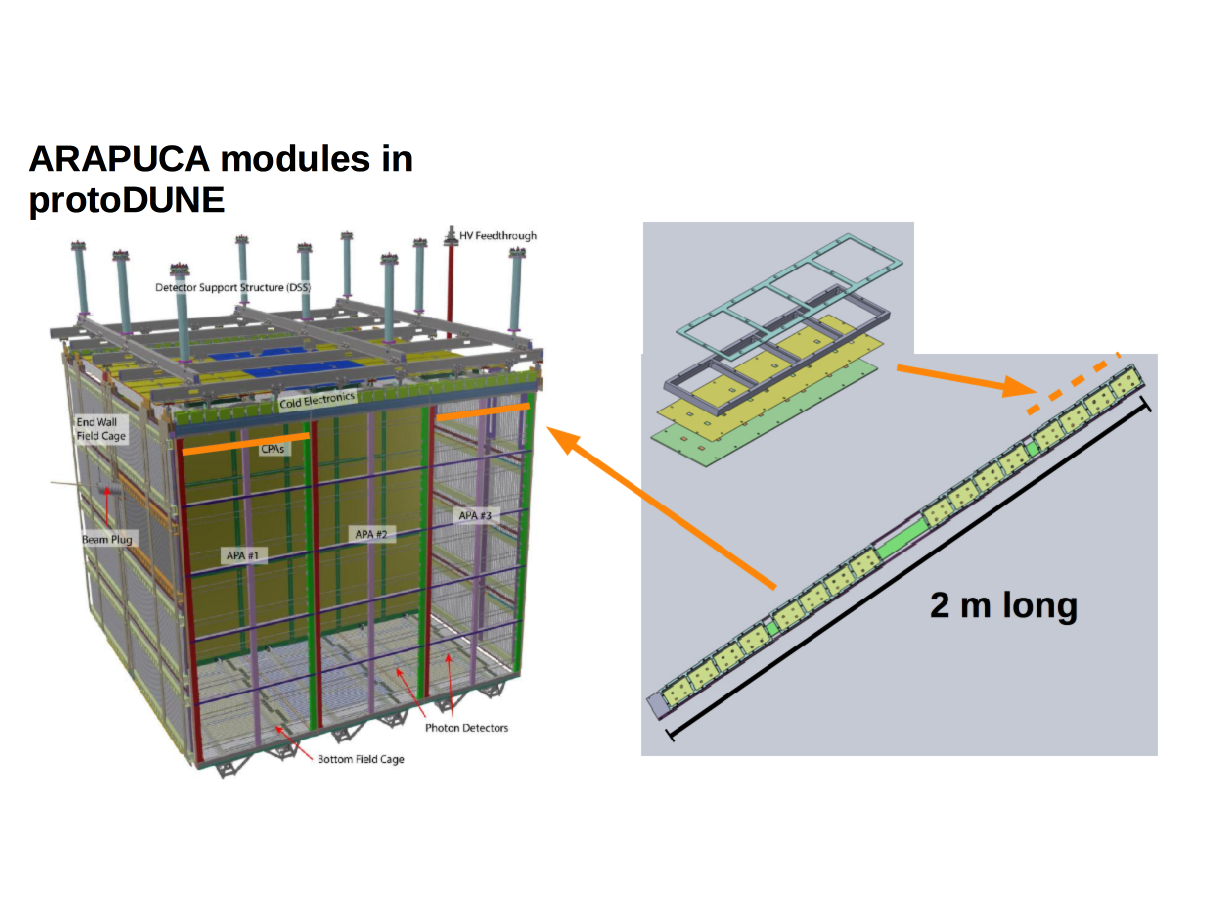}
\caption{ Left: The protoDUNE detector drawing showing the placement of two ARAPUCA modules (orange lines). The modules were designed to perfectly fit in the slots of the current PDS acrylic bars light guides \cite{IUbars}. Right: Detail of the ARAPUCA module (2 m long), each module hosts 4 {\it cartuchos}, which is an assembling of 4 ARAPUCA 10 $\times$ 8 cm$^2$ cells.}
  \label{Arap-pDUNE}
\end{figure}

The protoDUNE experimental program will test and validate the technologies and design that will be applied to the construction of the DUNE Far Detector at the Sanford Underground Research Facility (SURF) \cite{DUNE}. The protoDUNE detectors will be run in a dedicated beam line at the CERN SPS accelerator complex. The protoDUNE elements are \textbf{1:1 scale} as the current DUNE's design.
In protoDUNE we will equip the Photon Detection System (PDS) with 2 ARAPUCA modules as already agreed by the DUNE collaboration. Figure \ref{Arap-pDUNE} shows the ARAPUCA modules in the protoDUNE detector.
The ARAPUCAs for protoDUNE have modifications aiming to increase the overall performance. The SiPMs are soldered on the backplane, looking towards the internal side of the filter. This choice was driven by results from GEANT4 simulation, comparing the proposed design for protoDUNE with the previous TallBo configuration. Another possibility is to increase the SiPM number from 4 up to 12 and thus the ARAPUCA efficiency. The final number will be decided after the results from ongoing tests on the signal to noise ratio dependence on the number of SiPMs passively ganged.
            
\section{Conclusions}
\label{Conclu}

Noble liquids/gases detectors are the new paradigm in the field of neutrinos and dark matter research. Detection of scintillation light is a key piece on the detector performance. ARAPUCA is a promising device to be used as a photon detector in LArTPCs. We tested six different configurations. The average efficiency considering the six tested ARAPUCAs is $<\epsilon_{arap}>=0.37 \pm 0.01 (sys) \pm  0.03 (stat) \%$, which is comparable to other devices with same purposes \cite{IUbars}. No substantial differences can be quoted on the performance of individual components, despite one should note that by considering $A_{eff}$ and $\epsilon_{arap}$ side-by-side is possible to see that some amplification is achieved taking into account the ratio between sensitive area (the total area of SiPMs) and the acceptance window. However, it is worth mentioning that a R\&D program is on the way to improve the performance of the ARAPUCAs. Simulations to determine an optimal geometry are in course. Improvements on reflectors and WLS can be achieved. SiPMs ganging circuits (under development) can enlarge the sensitive area while keeping the number of readout channels.  We foresee the possibility to reach few \%  based only on the baseline design presented in this work and all the possible enhancements identified with the first experiments. A realistic scale test should be done in protoDUNE Photon Detection System (2018). Moreover, the same device can be used in Dark Matter experiments -- based on liquefied noble gases --  which requires large sensitive areas for photon detection. In this case the ARAPUCA concept can be adapted to increase the sensitive area in a cost-effective manner.

\acknowledgments
The authors thank Kenneth Treptow,  William Miner and Ronald P. Davis for their valuable knowledge and technical support. This work was supported by the Fermilab Neutrino Division through the NPC Fellowship program, and by the Brazilian agencies S\~{a}o Paulo Research Foundation (FAPESP), and National Council for Scientific and Technological Development (CNPq).

\paragraph{Note added.} It is worth to mention that another experiment made after the one reported in this paper shown $\epsilon_{arap} \sim 1.0\%$. Possible causes for the discrepancy are under analysis.


\end{document}